\documentclass[a4paper,twoside]{article}

\usepackage{epsfig}
\usepackage{subcaption}
\usepackage{calc}
\usepackage{amssymb}
\usepackage{amstext}
\usepackage{amsmath}
\usepackage{amsthm}
\usepackage{multicol}
\usepackage{pslatex}
\usepackage{apalike}
\usepackage{algorithm2e}
\usepackage[bottom]{footmisc}
\usepackage{SCITEPRESS}   

\usepackage{multirow} 
\usepackage{booktabs} 

\newcommand\blfootnote[1]{
    \begingroup
    \renewcommand\thefootnote{}\footnote{#1}
    \addtocounter{footnote}{-1}
    \endgroup
}

\begin{document}

\title{Integrating Traditional Technical Analysis with AI: A Multi-Agent LLM-Based Approach to Stock Market Forecasting}

\author{\authorname{Michał Wawer\orcidAuthor{0009-0004-2717-1616}, Jarosław A. Chudziak\orcidAuthor{0000-0003-4534-8652}}
\affiliation{Institute of Computer Science, Warsaw University of Technology, Warsaw, Poland}
\email{\{michal.wawer.stud, jaroslaw.chudziak\}@pw.edu.pl}
}

\keywords{Multi-Agent Systems, Elliott Wave Principle, Large language models (LLMs), Investment strategies, Deep Reinforcement Learning (DRL)}

\abstract{
Traditional technical analysis methods face limitations in accurately predicting trends in today's complex financial markets. This paper introduces ElliottAgents, an multi-agent system that integrates the Elliott Wave Principle with AI for stock market forecasting. The inherent complexity of financial markets, characterized by non-linear dynamics, noise, and susceptibility to unpredictable external factors, poses significant challenges for accurate prediction. To address these challenges, the system employs LLMs to enhance natural language understanding and decision-making capabilities within a multi-agent framework. By leveraging technologies such as Retrieval-Augmented Generation (RAG) and Deep Reinforcement Learning (DRL), ElliottAgents performs continuous, multi-faceted analysis of market data to identify wave patterns and predict future price movements. The research explores the system's ability to process historical stock data, recognize Elliott wave patterns, and generate actionable insights for traders. Experimental results, conducted on historical data from major U.S. companies, validate the system's effectiveness in pattern recognition and trend forecasting across various time frames. This paper contributes to the field of AI-driven financial analysis by demonstrating how traditional technical analysis methods can be effectively combined with modern AI approaches to create more reliable and interpretable market prediction systems.
}

\onecolumn \maketitle \normalsize \setcounter{footnote}{0} \vfill

\section{\uppercase{Introduction}}
\label{sec:introduction}

\blfootnote{\noindent \hspace{-0.65cm}\hrulefill\\ This is the accepted version of the paper presented at the \textbf{17th International Conference on Agents and Artificial Intelligence} (ICAART 2025), Porto, Portugal. Available at: https://doi.org/10.5220/0013191200003890}

The development of AI, including LLMs, has significantly increased interest in multi-agent systems \cite{agents_1,agents_2}. These advancements enable each agent to specialize in a specific area, enhancing the overall capability and performance of multi-agent systems beyond what was previously possible.

Traditional methods of predicting future stock prices using AI have often yielded unsatisfactory results due to limitations in processing vast amounts of data \cite{multiagent_stock_market_4,multiagent_stock_market_5} and adapting to rapidly changing market conditions.

How can a multi-agent system enhanced by LLMs improve the interpretability and efficiency of financial market trend analyses using the technical analysis method - Elliott Wave Principle (EWP)?
Utilizing framework for orchestrating AI agents, combined with advanced technologies like Retrieval-Augmented Generation (RAG) \cite{rag_2}, Deep Reinforcement Learning (DRL) and dynamic context management \cite{dynamic_context}, we have created ElliottAgents, a system designed to analyze stock market using LLM-based agents and EWP.

\begin{figure}[!t]
    \centering
    \includegraphics[width=7cm]{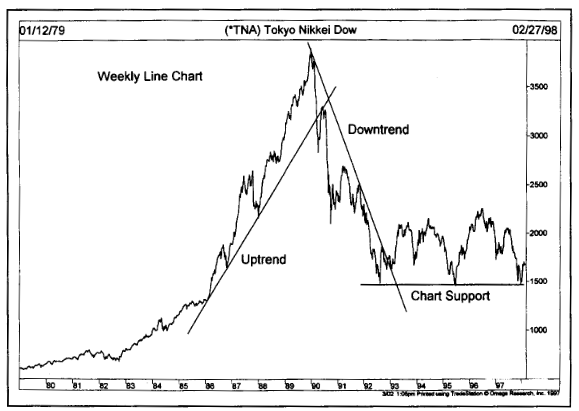}
    \caption{Basic trends in stock market, presented in \cite{elliott_theory_2}}
    \label{fig:murphy_basic_chart}
\end{figure}

The EWP is a form of technical analysis that investors use to forecast markets trends, which are presented on basic stock market chart on Fig. \ref{fig:murphy_basic_chart}. This theory identifies extremes in investor psychology, highs and lows in prices, and other collective factors by recognizing patterns described by Elliott Ralph Nelson \cite{elliott_theory_1}. By applying EWP through agents, it is possible to analyze these patterns more efficiently than traditional agent systems or manual analysis \cite{main_article}. Additionally, agents can learn from previous interactions and, over time, refine their strategies to determine what works best for a given company. This enables more accurate stock market forecasts, creating new opportunities for investors and analysts. By combining classical and contemporary approaches, we aim to create a platform for traders that can support them with multi-aspect analysis in the investment-making process.

The experiments demonstrate that agents are capable of recognizing wave patterns and analyzing them based on their knowledge, creating forecasts of future stock prices that can be utilized by traders and analysts. The results indicate that these agents can adapt to changing financial markets, achieving profits in the medium and long term. Furthermore use of DRL \cite{drl_2,drl_3} can improve these results by implementing continuous learning mechanism to constantly adapt to changing market conditions. However, their effectiveness in short-term is limited due to large minute-to-minute fluctuations and the presence of other high-frequency trading algorithms in the market.

\section{\uppercase{Background}}
\label{sec:realted-work}

\subsection{Prior work}
The integration of AI and multi-agent systems in stock market analysis has been a subject of ongoing research over the past two decades. An early attempt to combine traditional technical analysis with AI-driven approaches resulted in a multi-agent recommendation system \cite{main_article}. This system utilized a Java Agent Development (JADE) framework to implement a multi-agent architecture, demonstrating the potential of distributed analysis in financial forecasting. However, this early implementation was limited by the computational capabilities of its time and lacked the advanced natural language processing (NLP) capabilities now available.

The use of fuzzy logic within a multi-agent framework for technical analysis was explored in previous research \cite{multiagent_stock_market_4}. This approach introduced a flexibility in decision-making processes, allowing for better handling of market uncertainties. While innovative, the system's reliance on static rules and fuzzy logic constrained its adaptability to rapidly changing market conditions, a limitation that more recent AI technologies have sought to overcome.

Earlier work on multi-agent decision support systems for stock trading highlighted the potential of collaborative agent-based approaches in financial analysis \cite{multiagent_stock_market_5}. This research emphasized the importance of integrating diverse data sources and decision-making strategies within a multi-agent framework. However, the system's effectiveness was limited by the absence of advanced machine learning techniques that have since become integral to AI-driven financial analysis.

More recent advancements have seen the integration of neural networks and deep learning in stock market prediction systems \cite{szydlowski_chudziak_hidformer}. The potential of deep neural networks in enhancing profit through stock price prediction has been demonstrated in recent studies \cite{nn_2}. This work showcased the ability of neural networks to capture complex patterns in financial data, yet it did not fully address the interpretability challenges often associated with deep learning models in financial decision-making contexts.

\subsection{Research gap}

Despite advancements in the integration of AI and multi-agent systems, existing stock market analysis tools exhibit several limitations.

Firstly, many AI-driven financial forecasting tools, such as those leveraging neural networks, excel in recognizing complex patterns but fail to provide interpretable results. This lack of interpretability undermines trader trust and limits actionable insights. Secondly, early implementations, such as fuzzy logic-based systems \cite{multiagent_stock_market_4}, introduced flexibility in decision-making but were hindered by static rule sets that could not adapt to rapidly changing market conditions. Thirdly, while some studies \cite{main_article} attempted to incorporate the EWP into multi-agent frameworks, these efforts were limited by outdated computational capabilities and lacked advanced tools for automated wave pattern recognition.
Furthermore, existing tools often focus exclusively on either technical analysis or purely data-driven methods. 

To address these gaps, we introduce ElliottAgents system \cite{paclic_elliottagents}. By integrating the EWP into its core framework, ElliottAgents ensures that market analyses are grounded in well-established financial theories, making predictions more transparent and interpretable for traders. The system employs LLMs for improved natural language understanding and decision-making, and RAG to access external knowledge bases, ensuring up-to-date and contextually relevant analyses.

In addition, ElliottAgents leverages DRL to incorporate a continuous learning mechanism that refines strategies based on historical data, allowing it to adapt to evolving market conditions. The system’s architecture also enables specialized agents to collaborate dynamically, each focusing on distinct tasks such as data processing, pattern recognition, and strategy formulation. This collaborative approach ensures efficiency and scalability.

\section{\uppercase{Theoretical Foundations}}
\label{sec:theoretical-foundations}

\subsection{Elliott Wave Principle (EWP)}

\begin{figure}[!b]
    \centering
    \includegraphics[width=2.5in]{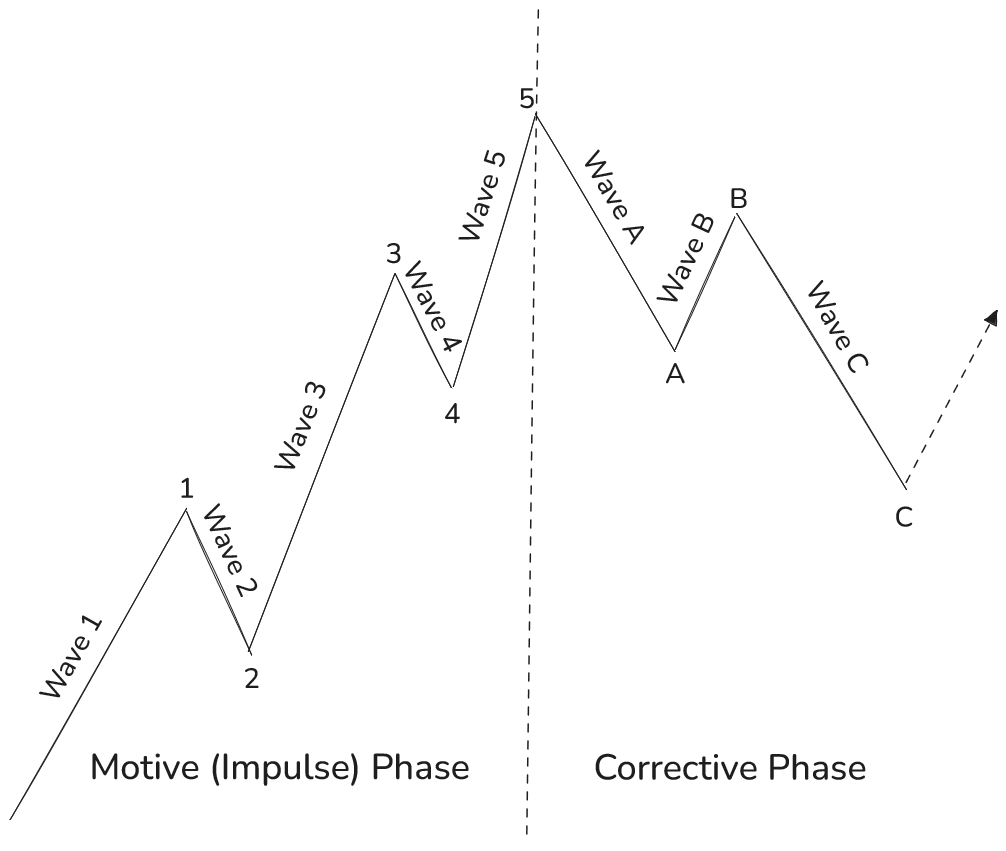}
    \caption{Impulse and corrective waves patterns, adapted from Prechter and Frost, 1978 \cite{elliott_theory_1}}
    \label{fig:elliott_example}
\end{figure}

The EWP, introduced by Ralph Nelson Elliott, is a technical analysis framework that suggests market prices follow identifiable patterns influenced by collective investor behavior and psychology \cite{elliott_theory_1,elliott_theory_2}. According to this principle, market trends alternate between periods of optimism and pessimism, producing consistent wave-like price movements. Elliott categorized these patterns into thirteen recurring structures, referred to as "waves" which are broadly divided into two primary types: impulsive waves and corrective waves.

 Motive (impulsive) waves, presented on Fig. \ref{fig:elliott_example}, are the driving force behind market trends and consist of five sub-waves. These five sub-waves move in the direction of the overall trend. Within an impulsive wave, waves 1, 3, and 5 are the main movement waves, while waves 2 and 4 are corrective and move against the trend. The structure of an impulsive wave ensures progress in the direction of the primary trend, with Wave 3 typically being the strongest and longest of the three impulsive waves. Corrective waves, move against the main trend and consist of three sub-waves labeled A, B, and C. These corrective waves provide a counterbalance to the impulsive waves, retracing a portion of the preceding trend.

Market movements can be broken down into larger and smaller waves, creating a fractal-like structure named wave degrees. Smaller waves combine to form larger waves, which in turn combine to form even larger waves, creating a nested pattern. This fractal nature allows the EWP to be applied to different time frames, from short-term market movements to long-term trends.

The EWP does not offer certainty but provides a framework for assessing the probabilities of different market scenarios. It helps traders understand the current market context and predict potential future paths, making it a valuable tool for technical analysis. 

\subsection{Fibonacci approach in EWP}

The Fibonacci sequence is integral to the EWP, providing a mathematical framework that enhances the predictability and structure of market movements. The Fibonacci sequence is a series of numbers where each number is the sum of the two preceding ones. This sequence is known for its prevalence in nature, art, and architecture, and it similarly manifests in the financial markets.

\begin{figure}[!b]
    \centering
    \includegraphics[width=2.5in]{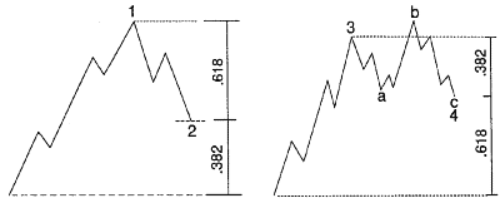}
    \caption{Fibonacci retracements in corrective waves \cite{elliott_theory_1}}
    \label{fig:fibo_retracements}
\end{figure}

In the context of the EWP, the Fibonacci sequence helps to quantify the relationships between different waves in the market \cite{fibonacci_theory_1}. Elliott observed that market waves often unfold in a pattern that aligns with Fibonacci ratios \cite{elliott_theory_1}. Based on Fig. \ref{fig:fibo_retracements}, the length of one wave might be 1.618 times the length of another, reflecting the Golden Ratio, which is approximately 1.618. This ratio, also known as Phi (\(\varphi\)), is fundamental to the proportionality observed in wave patterns.

The fractal nature of Elliott waves means that Fibonacci relationships apply across different degrees of trend, from minute charts to long-term market cycles \cite{fibonacci_theory_1}. This fractal characteristic ensures that patterns observed on smaller scales can be seen on larger scales, maintaining the same Fibonacci proportionality. For instance, a complete market cycle might consist of a series of waves that adhere to Fibonacci ratios, creating a cohesive and predictable structure throughout the market's evolution.

\subsection{LLMs in time series prediction}

Time series prediction using LLMs has emerged as a powerful approach in financial forecasting, integrating the latest advancements in AI to improve the accuracy and reliability of predictions. At its core, time series analysis aims to understand the underlying structure and function that produce the observed data. This understanding is then used for forecasting future values of the series.Traditional time series analysis methods, such as ARIMA and exponential smoothing, focus on decomposing a series into trend, seasonal, and residual components to identify patterns and predict future values \cite{time_series_1}. These models rely on historical data and linear assumptions, often struggling with non-linear and complex temporal dependencies inherent in financial markets.

LLMs, on the other hand, leverage deep learning techniques and extensive datasets to understand and predict time series data. These models leverage their capability to understand and generate sequential data, which is crucial for accurate forecasting of time series characterized by trends and seasonal patterns. 

Methods such as natural language paraphrasing and incorporating external knowledge into prompts have been demonstrated to enhance their performance further \cite{time_series_llm_2}. However, challenges remain, particularly with multi-period datasets where LLMs struggle to recognize distinct periods \cite{time_series_llm_3}, similar problems apply to all other methods \cite{nn_1}. Despite their computational demands, LLMs often perform on par with simpler models, suggesting that they hold potential but more research is needed to prove their effectiveness. The use of agents may be a factor that will greatly improve the results of time series prediction by distributing tasks among agents, enabling a more robust analysis of complex big sets of data.

\subsection{Deep Reinforcement Learning (DRL)}

As a subset of machine learning, called DRL integrates principles of deep learning and reinforcement learning (RL). In RL, an agent learns to make sequential decisions by interacting with an environment to maximize cumulative rewards. This process involves observing the current state, selecting actions, and receiving feedback in the form of rewards, iterating this cycle to improve the agent’s policy, which is the strategy for choosing actions \cite{drl_1}.

Deep learning, which uses neural networks with multiple layers, enhances RL by enabling the handling of high-dimensional state and action spaces. Key DRL algorithms include Deep Q-Networks (DQN), which use neural networks to estimate Q-values (expected rewards for actions), and Policy Gradient methods, which directly optimize the policy \cite{drl_2}. DRL leverages techniques like experience replay, where past experiences are stored and reused during training, and target networks, which help stabilize training by providing consistent target values.

In the backtesting process, we use DRL to analyze historical market data \cite{drl_3}. A DRL agent can learn which patterns are effective for a given company and understand how each pattern can affect future price movements. This enables the agent to make informed buy, sell, or hold decisions, optimizing long-term returns. By continuously learning and adapting, DRL agent will increase accuracy in dynamic and uncertain environments.

\subsection{Multi-agent architecture}

Multi-agent systems (MAS) have a longstanding role in modeling complex systems, where autonomous agents interact with each other and their environment \cite{agents_11}. These systems were traditionally built using rule-based systems, symbolic equations, stochastic modeling, and early machine learning techniques \cite{agents_12}. 

The integration of LLMs, such as ChatGPT, has significantly enhanced MAS by equipping agents with advanced NLP capabilities \cite{agents_8}. NLP enables agents to comprehend complex instructions, collaborate effectively, and explain their actions, thereby increasing transparency and trust within the MAS. LLMs allow agents to operate more autonomously, dynamically perceiving and responding to changes in their environment while learning from experiences to adapt to new situations without explicit instructions \cite{agents_1}. This learning process mirrors human behavior, allowing for more realistic simulations. 

LangChain is a framework that facilitates the chaining of different components within an LLM application, including agents \cite{langchain_1}. Our system utilizes LangGraph, a LangChain component, to visualize and manage relationships between agents, enhancing clarity and interpretability in complex multi-agent interactions. Agents in an LLM-powered MAS collaborate, performing sequential and hierarchical tasks that culminate in a comprehensive analysis. Some agents utilize advanced tools that enhance their analytical capabilities, allowing for the generation of more precise and accurate results.

\subsection{ReAct agent}

The ReAct paradigm \cite{react_agent} represents a significant advancement in leveraging LLMs for complex problem-solving tasks. ReAct, which stands for "Reasoning + Acting," combines the strengths of chain-of-thought reasoning (series of intermediate steps to arrive at a solution) with the ability to interact with external environments, creating a more robust and adaptable system for tackling diverse challenges.

At its core, ReAct prompts LLMs to generate both verbal reasoning traces and task-specific actions in an interleaved manner \cite{react_agent}. This approach allows the model to perform dynamic reasoning to create, maintain, and adjust high-level plans for acting (reason to act), while also interacting with external environments to incorporate additional information into its reasoning process (act to reason).

\section{\uppercase{ElliottAgents System Architecture}}
\label{sec:ellioagents-system-architecture}

The development of ElliottAgents aims to integrate traditional financial analysis with modern AI capabilities. This section outlines the basic assumptions and design principles underlying the platform's architecture and implementation.

\subsection{ElliottAgents design approaches}

The increasing complexity of financial markets, coupled with recent advances in AI, presents both opportunities and challenges for developing new market analysis tools. ElliottAgents system's architecture is designed to support these functions:
\begin{enumerate}
  \item \textbf{Configurable analysis parameters}: 
  The platform implements a flexible parametrization that allow user to specify the asset, timeframe, and data granularity.
  
  \item \textbf{Dynamic Data Integration}: ElliottAgents incorporates real-time market data through an external yfinance API, enabling analysis of large collections, the most recent data.
    
  \item \textbf{Pattern recognition and analysis}: The platform implements algorithms for identifying Elliott Wave patterns across multiple timeframes as a tool for agents. Then results of this tool are interpreted by LLM-based agents. The integration of AI enables more nuanced pattern recognition than traditional technical analysis methods.
  
  \item \textbf{Multi-Agent collaboration}: A crew of specialized agents works in concert to analyze market data. Each agent maintains specific expertise, from data processing to pattern recognition and strategy formulation. The collaborative framework enables comprehensive market analysis through the synthesis of multiple analytical perspectives.
  
  \item \textbf{Continuous learning and optimization}: The platform incorporates continuous learning mechanisms that enable it to refine its analytical capabilities over time. This includes real-time data processing, strategy backtesting, and performance optimization through DRL implementation.
\end{enumerate}

\subsection{Agents definition}

\begin{figure}[!t]
    \centering
    \includegraphics[width=7cm]{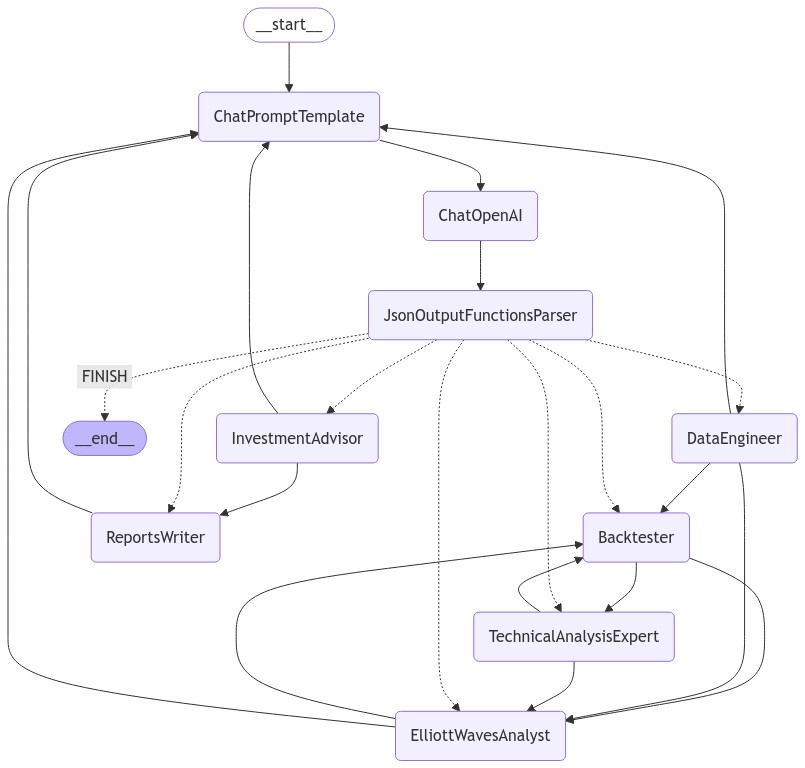}
    \caption{Graph presenting data flow between agents, generated in LangGraph framework.}
    \label{fig:agents_hierarchical}
\end{figure}

At the core of ElliottAgents is a multi-agent architecture that orchestrates specialized agents, each responsible for distinct aspects of the analysis process. These agents can dynamically perceive and respond to changes in their environment, learning from their experiences to improve future responses. The architecture of ElliottAgents consists of 7 agents who communicate with each other in a structured way, as shown in the Fig. \ref{fig:agents_hierarchical}.

The Coordinator agent orchestrates the entire process. It begins by receiving user input, including the desired stock symbol and analysis timeframe. This information is then passed to the Data Engineer, which gathers the necessary historical stock data. Next, the Coordinator triggers the Elliott Waves Analyst, equipped with a specialized tool to identify and classify Elliott Wave patterns within the historical data. These patterns are visually represented through automatically generated charts.
The Backtester agent then receives the identified patterns and employs DRL to validate them against historical trends, assessing their effectiveness. The Technical Analysis Expert then steps in, combining the Elliott Wave patterns with the backtesting results. This agent determines the most probable pattern for the current market conditions, providing a refined prediction. This refined prediction, along with the original data and wave patterns, is then forwarded to the Investment Advisor. This agent synthesizes all the information, incorporating insights from a RAG tool, to formulate a comprehensive investment strategy. This strategy includes specific buy/sell signals, price targets, and contingency plans.
Finally, the Reports Writer receives all the compiled information and generates a clear, concise report for the end-user. This report presents the investment strategy in an easily understandable format, ensuring the user has actionable insights based on the latest data and analysis.

Each agent in the system is provided with a specific context through a natural language prompt, enabling it to perform its designated role effectively. The following prompt defines the context and responsibilities of the Investment Advisor agent, focusing on synthesizing analyses into actionable investment strategies.

\begin{small}
\begin{verbatim}
You are the Investment Advisor. Your role is 
crucial in synthesizing the analyses provided 
by other agents and formulating actionable 
investment advice. Your tasks include:
1. Interpret the Elliott Wave patterns 
identified by the Elliott Waves Analyst.
2. Consider the backtesting results provided 
by the Backtester agent.
3. Integrate information from the rag tool
to provide context to your recommendations.
4. Formulate a comprehensive investment strategy 
that includes:
- Buy, sell, or hold recommendations
- Price targets for entry and exit points
- Stop-loss levels
- Time frames for the transactions
5. Highlight any potential risks or limitations 
in the analysis, ensuring a balanced view of 
the investment opportunity.
6. Provide any additional insights that could 
be valuable for decision-making, such as 
correlations with broader market trends.

Remember, your advice should be based on the 
collective intelligence of the multi-agent 
system. Aim to present your investment advice 
in a structured format that can be easily 
understood the end users.
\end{verbatim}
\end{small}

\subsection{Agents customization}

\begin{figure}[!b]
    \centering
    \includegraphics[width=7cm]{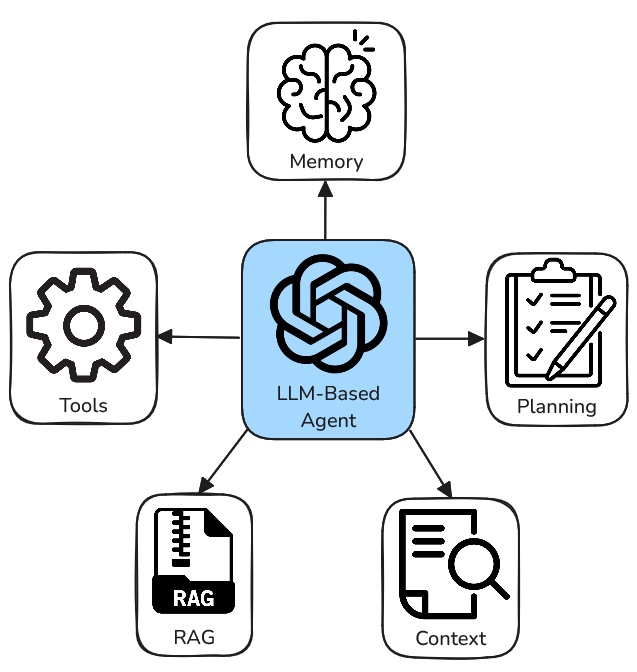}
    \caption{Overview of a LLM autonomous agent.}
    \label{fig:agent}
\end{figure}

As presented on Fig. \ref{fig:agent} agent is build using different components, technologies used by our agents are described below: 
\begin{itemize}
    \item \textbf{Memory} preserve and regulate knowledge, experiential data, and historical information \cite{agents_5}. It typically consists of short-term memory (for immediate context and task-related information) and long-term memory (for storing substantial volumes of knowledge and past experiences). Memory mechanism helps agents generate responses based on past interactions, improving decision-making and context-awareness over time \cite{agents_2}.
    
    \item \textbf{Planning} is a ability to devise action sequences based on set objectives and environmental constraints \cite{agents_4}. For LLM-based agents, planning often utilizes in-context learning methods like Chain of Thought (CoT), Tree of Thought (ToT), or external capabilities. It involves task analysis, action anticipation, and optimal action selection to address complex problems.
    
    \item \textbf{Context} refers to the ability of AI agents to adaptively adjust their contextual understanding based on real-time information \cite{dynamic_context}. Agents can utilize various types of context, including tools, documents accessed through RAG, the history of conversations, and the ability to reflect and plan future actions. This approach leverages ongoing interactions and updates the context dynamically, enabling the agent to maintain relevance and accuracy throughout a session. By incorporating new data as it becomes available, dynamic context helps agents refine their responses and improve decision-making processes. 
    
    \item \textbf{Retrieval-Augmented Generation (RAG)} improves the factual reliability of generative AI by integrating external information retrieval into its workflow \cite{rag_2}. Instead of relying solely on pre-trained parameters, RAG dynamically fetches relevant data from external knowledge bases, which enhances both the accuracy and relevance of generated responses. This process involves encoding user queries as embeddings, comparing them against a vectorized database, and incorporating the retrieved information into the model's output \cite{rag_3}. By reducing hallucinations and providing traceable sources, RAG addresses common challenges in AI-driven content generation.
    
    In our system, a knowledge graph-based RAG framework is employed. This structure organizes data into interconnected graphs, allowing for precise query disambiguation and improved contextual relevance. Leveraging these structured relationships, our approach supports advanced tasks like interpreting the mathematical underpinnings of the EWP while ensuring accuracy and consistency in the generated responses.
    
    \item \textbf{Tools} LLM-based agents often integrate various tools to enhance their problem-solving abilities and interact with external environments or data sources. The tools component presents specific instruments that were created for our system to adjust it for analysis of financial markets:
    \begin{itemize}
        \item \textbf{Stock market data}: provides access to real-time and historical market information, essential for informed decision-making.
        \item \textbf{Elliot waves reader}: tool for technical analysis, helping the agent identify and interpret Elliot wave patterns in price movements.
        \item \textbf{Chart generator}: allows the agent to visualize market data, especially Elliott waves, creating graphical representations of detected waves. This tool is directly connected to Elliott waves reader tool.
        \item \textbf{Database connector}: enables the agent to access and manage structured data of backtesting results.
    \end{itemize}
\end{itemize}

\subsection{Continous learning agent}

The ElliottAgents platform implements a continuous learning process \cite{continous_learning} that enables agents to adapt and refine their knowledge over time \cite{agents_10}. This process is designed to enhance the system's predictive capabilities without relying on traditional fine-tuning methods. Instead, agents learn organically through their interactions and observations of the stock market environment.

At the core of this process is the Backtester agent, which plays a crucial role in accumulating and leveraging historical knowledge. The Backtester's workflow begins with a query to determine if relevant results are already available in the backtesting knowledge base. If not, the agent initiates a analysis by fetching the necessary data, performing EWP analysis, and interpreting the results. These findings are then stored in the Neo4j graph database for future reference. This iterative process allows the system to build a repository of analyzed patterns and outcomes over time. This approach ensures that the system's predictions are grounded in a historical data and previously observed market behaviors.

\subsection{Agents flow engineering}

\begin{figure}[!t]
    \centering
    \includegraphics[width=7.5cm]{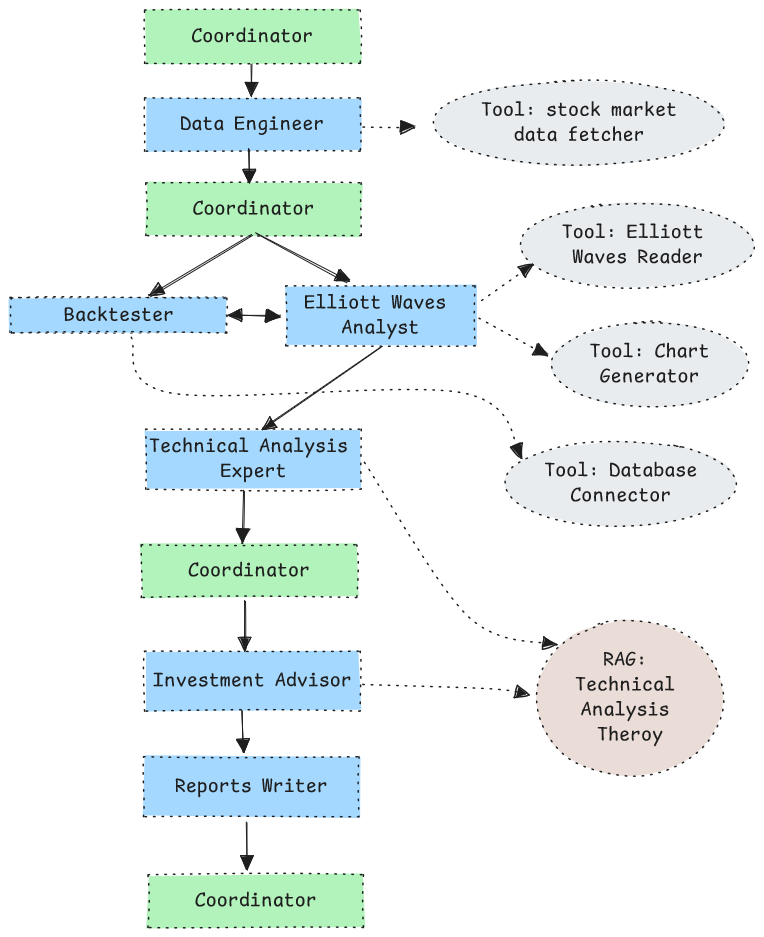}
    \caption{Interactions between agents and tools.}
    \label{fig:experiment_flow}
\end{figure}

The orchestration of the system is designed to facilitate seamless collaboration among agents, ensuring an efficient workflow. The system employs a hierarchical structure wherein each agent is assigned a specialized role \cite{agents_8}, contributing to scalability. The coordinator is the agent who manages the whole flow of information in the system, distributes the tasks and ensures their execution, as presented on Fig. \ref{fig:experiment_flow}.

Asynchronous task execution enables agents to operate in parallel, mitigating bottlenecks and enhancing throughput \cite{agents_5}. Tasks such as backtesting and wave analysis, which do not require immediate interdependence, are executed concurrently. This asynchronous design is further supported by the system’s capacity for dynamic scaling, instantiating additional agents when computational demands increase.

Task decomposition and memory management are fundamental to the system's architecture \cite{agents_9}. Complex tasks are divided into smaller, manageable subtasks, allowing agents to focus on well-defined objectives. Memory management ensures continuity through memory identifiers, enabling agents to retain context across tasks. For example, outputs from the Elliott Wave Analyst inform subsequent validation by the Technical Analysis Expert, leading to actionable strategies devised by the Investment Advisor and finalized in reports by the Reports Writer under the coordinator's supervision.

The designed flow is integral to its efficiency, scalability, and resilience. By delegating tasks to specialized agents and leveraging dynamic scaling, the system adapts to varying demands without compromising performance. The asynchronous execution reduces latency, while the collaborative interactions between agents ensure the production of accurate outputs \cite{agents_7}. These design principles collectively enable the system to perform complex analytical tasks with precision, speed, and adaptability.

\section{\uppercase{Experimental Setup and Methodology}}
\label{sec:experimental-setup}

The experiments conducted on the created platform were designed to test its use in the real market, effectiveness of pattern recognition and to study the impact of the backtesting process on the final result. Historical The data on which we conducted our tests came from the yfinance library, we focused on 2 time intervals: daily and hourly. In the performed tests, our agents used the gpt-4o-mini model from OpenAI. Currently, ElliottAgents allows the detection of only a few wave patterns, there are more patterns discovered and described in Elliott Wave Theory, and in our study we focused on describing only a few selected ones. Current state of our system is able to recognize impulse and corrective patterns, with additional wave extensions. By recognizing these patterns we are able to determine support, resistance and target price levels.

The experimental evaluation of the ElliottAgents platform was conducted in two distinct phases, each designed to assess different aspects of the system's performance and reliability in stock market prediction.

The initial phase of our experimentation focused on demonstrating the practical application of the ElliottAgents system through a detailed case study. We selected a specific company and time frame to showcase a complete analysis cycle. This phase aimed to illustrate the inter-agent communication process, highlighting how different specialized agents collaborate to produce a comprehensive market analysis. It also demonstrated the step-by-step reasoning and decision-making process employed by the agents. Furthermore, this analysis provided insights into the potential real-world applicability of the system's outputs, including the identification of Elliott Wave patterns and the generation of trading signals.

The second phase of our experimentation was designed to assess the system's pattern recognition capabilities and the accuracy of its predictions. This evaluation utilized a cross-validation method applied to a substantial dataset of historical market price movements. We utilized 1,000 samples each representing a one candlestick from the price charts of selected stocks in daily and hourly interval. 

Our tests focused on two key Elliott Wave formations: incomplete impulsive waves (1-2-3-4) consisting of four sub-waves, and complete impulsive waves (1-2-3-4-5) comprising all five sub-waves. In both cases, we enforced the EWP rule that waves must not overlap. To evaluate the effect of our DRL component, we conducted parallel tests with and without the DRL backtesting process. The DRL model was trained on 10 years of historical data for each company under examination.

For each identified pattern, the system generated a prediction for the subsequent price movement (upward or downward). We evaluated these predictions using specific criteria. For incomplete waves (1-2-3-4), a prediction was deemed correct if the average price over the next n candlesticks (where n is calculated to approximate 1.62 times the length of the first wave) moved in the predicted direction. For complete waves (1-2-3-4-5), a prediction was considered accurate if the subsequent corrective wave A exhibited a exact same length to that of wave 5, moving in the opposite direction.

\section{\uppercase{Results and Analysis}}
\label{sec:results}

The first phase of the experiments was performed on Apple (AAPL) price data from September 2023, to September 2024. The system was run on limited historical data from to recognize all waves pattern and identify possible buy or sell signals. Based on them, we could simulate transactions and calculate theoretical investment returns, proving the effectiveness of predictions of future prices. Here we'll describe just one example of the analysis by presenting the message exchange between agents and the final analysis with the result of the simulated transaction.

\begin{figure}[!t]
    \centering
    \includegraphics[width=7.5cm]{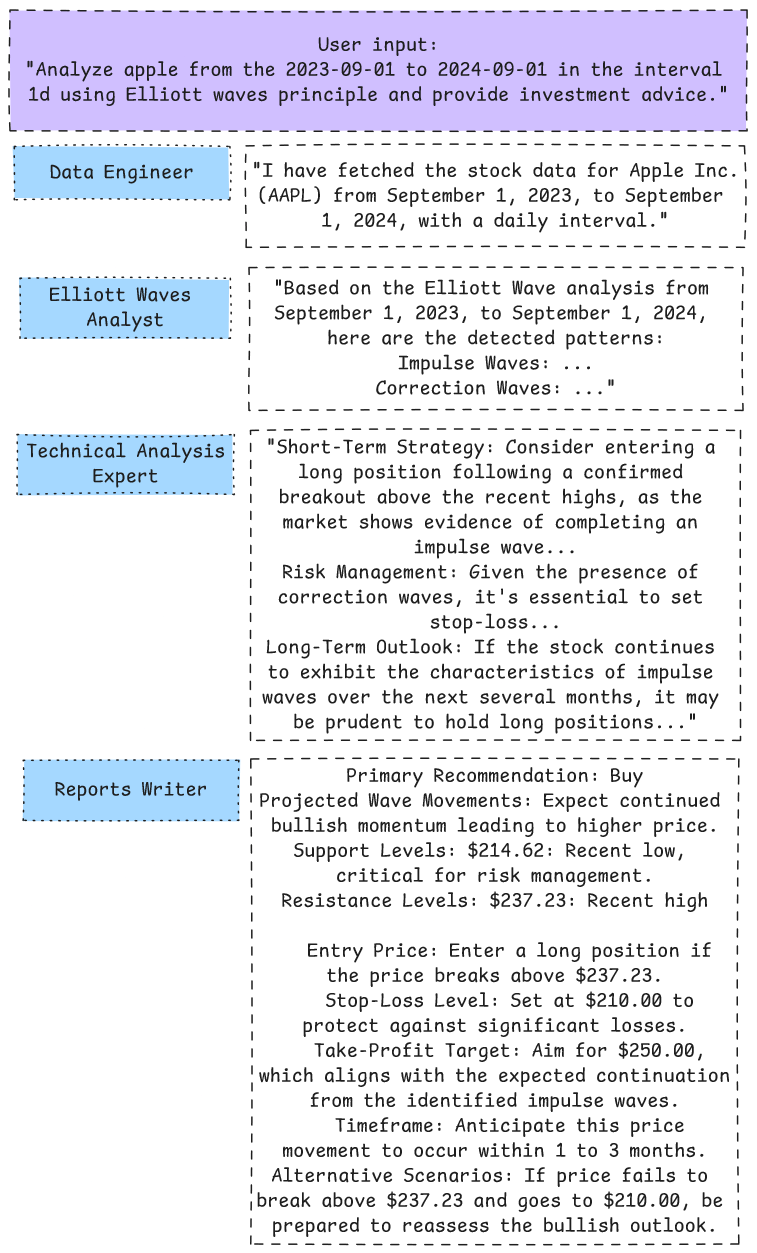}
    \caption{Example interactions between agents analyzing Apple stock, with messages returned by each agent.}
    \label{fig:experiment_chat}
\end{figure}

\begin{figure}[!b]
    \centering
    \includegraphics[width=7.2cm]{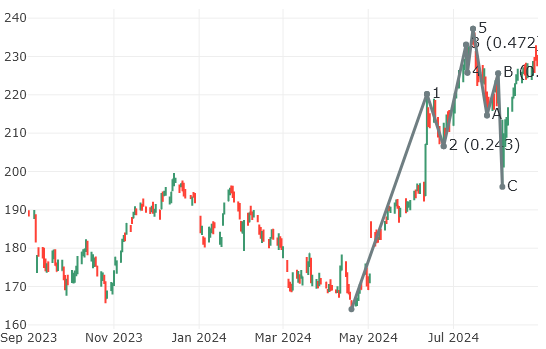}
    \caption{Impulsive and corrective waves found on Apple stock on 1d interval.}
    \label{fig:apple_1y_1d}
\end{figure}

The interaction between agents is depicted in Fig.~\ref{fig:experiment_chat}, while Fig.~\ref{fig:apple_1y_1d} illustrates a chart with analysis of Apple stock over a one-year period, with data aggregated at a daily interval. During this period, ElliottAgents successfully identified an impulsive wave sequence labeled 1-2-3-4-5 and a subsequent corrective wave pattern denoted as A-B-C. According to established wave theory, the occurrence of this configuration indicates a likely reversal surpassing the peak of the fifth wave.

Upon identifying this wave structure and confirming the initiation of a reversal, ElliottAgents issued a buy recommendation at a price of \$232 per share. The target price was set at \$250 per share, corresponding to the peak of the fifth wave. This price also considered the resistance level observed at the peak of wave B (\$225), resulting in a dual-target strategy. Such a strategy aims to ensure an optimal exit point while providing a buffer for potential resistance at critical levels. As illustrated in the chart, the prediction proved accurate, with the stock price stabilizing near the \$250 mark—aligned with the peak of the fifth wave—where it encountered notable resistance.

The second part of the experiment focus on quantity tests for the correctness of the detected pattern and the impact of DRL on results. Test was conducted using a cross-validation method on 1000 samples (candlesticks) with a daily interval for the stocks. We compared the results with and without a DRL backtesting process.

\begin{table*}
    \centering
    \begin{tabular}{l | c c c | c c c}
        \hline
        \multirow{2}{*}{\textbf{Stock}} & \multicolumn{3}{c}{\textbf{1-2-3-4 Patterns}} & \multicolumn{3}{| c}{\textbf{1-2-3-4-5 Patterns}} \\
        \cline{2-7}
         & {\bfseries N} & \begin{tabular}[c]{@{}c@{}}{\bfseries Without}\\{\bfseries backtesting}\end{tabular} & \begin{tabular}[c]{@{}c@{}}{\bfseries With}\\{\bfseries backtesting}\end{tabular} & {\bfseries N} & \begin{tabular}[c]{@{}c@{}}{\bfseries Without}\\{\bfseries backtesting}\end{tabular} & \begin{tabular}[c]{@{}c@{}}{\bfseries With}\\{\bfseries backtesting}\end{tabular} \\
        \hline
        \multicolumn{7}{l}{\textit{Daily Interval}} \\
        \hline
        \textbf{AMZN} & 24 & 58.34\% & 66.67\% & 18 & 66.67\% & 77.78\% \\
        \hline
        \textbf{GOOG} & 28 & 53.57\% & 67.86\% & 23 & 65.22\% & 82.61\% \\
        \hline
        \textbf{INTC} & 19 & 57.89\% & 73.68\% & 15 & 60.00\% & 73.34\% \\
        \hline
        \textbf{CSCO} & 12 & 58.34\% & 66.67\% & 9 & 66.67\% & 88.89\% \\
        \hline
        \textbf{ADBE} & 20 & 60.00\% & 65.00\% & 12 & 50.00\% & 66.67\% \\
        \hline
        \textbf{META} & 22 & 59.09\% & 63.64\% & 14 & 57.14\% & 64.29\% \\
        \hline
        \multicolumn{7}{l}{\textit{Hourly Interval}} \\
        \hline
        \textbf{AMZN} & 10 & 50.00\% & 70.00\% & 8 & 62.50\% & 75.00\% \\
        \hline
        \textbf{GOOG} & 13 & 53.84\% & 61.54\% & 9 & 77.78\% & 77.78\% \\
        \hline
        \textbf{INTC} & 12 & 58.34\% & 66.67\% & 9 & 66.67\% & 88.89\% \\
        \hline
        \textbf{CSCO} & 9 & 44.45\% & 55.56\% & 8 & 50.00\% & 50.00\% \\
        \hline
        \textbf{ADBE} & 10 & 50.00\% & 70.00\% & 8 & 62.50\% & 75.00\% \\
        \hline
        \textbf{META} & 12 & 58.34\% & 58.34\% & 9 & 66.67\% & 88.89\% \\
        \hline
        \bottomrule
        \multicolumn{7}{l}{\footnotesize N: number of patterns found.}
    \end{tabular}
    \caption{Comparison of pattern recognition with and without backtesting}
    \label{tab:pattern_comparison}
\end{table*} 

Table \ref{tab:pattern_comparison} presents the results of the cross-validation experiments for 1000 data samples in two time intervals. As we can see, the identification of a complete impulsive wave pattern contributes to better predictions of subsequent price movements than incomplete impulse wave pattern. In case of hourly intervals our system detected smaller number of patterns, mainly because price changes on the hourly interval were smaller. The use of DRL resulted in a improvement in prediction, showing that agents are able to use the learning process on historical data in better interpretation of patterns.

\section{\uppercase{Discussion}}
\label{sec:discussion}

Multi-agent architectures have been utilized in stock price prediction systems for many years~\cite{multiagent_stock_market_1,multiagent_stock_market_4,multiagent_stock_market_5}. However, advancements in AI over recent years have significantly enhanced the capabilities of these systems. The introduction of fuzzy logic in earlier systems provided a foundation for integrating qualitative judgments with quantitative analysis. Nevertheless, these systems required further optimization to improve their decision-making processes. While it remains challenging to directly compare the profitability of our system with other stock price prediction systems currently available, experimental results indicate that our approach effectively detects and interprets wave patterns with greater accuracy than comparable systems utilizing EWP \cite{main_article}. Furthermore, the analyses generated by our agents present a clear investment plan, including actionable price levels, which can be directly applied by traders in real-world scenarios.

The experiments conducted on ElliottAgents have yielded several insights:
\begin{enumerate}
    \item \textbf{Pattern recognition accuracy}: The system demonstrated a high accuracy in identifying impulsive and corrective waves patterns across various time frames. Experiments conducted on historical stock market data validate the system's capability to recognize and interpret intricate market structures effectively.
    \item \textbf{Impact of backtesting}: The implementation of DRL for backtesting significantly enhanced the system's predictive accuracy. Across different companies and time intervals, backtesting improved pattern recognition validity by up to 16\%, demonstrating the importance of historical data analysis in refining predictive models.
    \item \textbf{Multi-Agent architecture effectiveness}: The distributed approach of ElliottAgents, where specialized agents handle different aspects of analysis, proved highly effective. This architecture allowed for more efficient processing of complex data and improved the overall accuracy of predictions.
\end{enumerate}

The system's ability to dynamically update context and integrate EWP, significantly improves the accuracy and reliability of the predictions. Backtesting capabilities usind DRL further allow for the continuous refinement of strategies.

The development and testing of ElliottAgents have successfully addressed the primary research question posed at the outset of this study. The platform has demonstrated that it is indeed possible to integrate the EWP into a multi-agent architecture to more quickly and accurately predict future stock price movements. 

The development and testing of ElliottAgents have successfully addressed the primary research question posed at the outset of this study. The platform has demonstrated that it is indeed possible to integrate the EWP into a multi-agent architecture to more quickly and accurately predict future stock price movements. By leveraging AI technologies, the system enhances both interpretability and efficiency, addressing the limitations of traditional methods. The collaborative multi-agent design ensures scalability and adaptability, making ElliottAgents a robust tool for modern financial analysis. Furthermore, the research has made substantial progress on several key objectives:
\begin{enumerate}
    \item \textbf{Multi-faceted analysis}: ElliottAgents have shown the ability to perform comprehensive analyses and present results in a user-friendly manner, making complex financial data accessible to both professional traders and individual investors.
    \item \textbf{LLMs in time series prediction}: The research has provided valuable insights into the performance of LLMs in time series prediction, particularly in the context of stock market trends. While challenges remain, the integration of LLMs with traditional technical analysis methods has shown promising results.
    \item \textbf{Real-time data integration}: The system has demonstrated the ability to effectively utilize the most recent stock market data, adapting to rapidly changing market conditions in near real-time.
    \item \textbf{Agent customization}: The use of advanced technologies such as RAG, and memory management techniques has allowed for better customization of agents for specific tasks, enhancing the overall performance of the system.
    \item \textbf{Multi-Agent cooperation}: The research has shown that the multi-agent approach improves performance compared to single-agent systems, particularly in complex market scenarios.
\end{enumerate}

\section{\uppercase{Future Work and Conclusion}}
\label{sec:conclusion}

\subsection{Future work}

Currently, our work has focused primarily on only few patterns recognized by EWP. Expanding platform to include additional wave formations such as truncations, zigzags, flat corrections, triangles, and other patterns could improve our predictive capabilities \cite{elliott_theory_1}. Following the successful integration of EWP, we could further improve our system by incorporating other technical analysis methods \cite{elliott_theory_2}, such as moving averages. This expansion could enhance our ability to determine more accurate buy or sell signals, potentially improving signal reliability and profitability. Multi-agent architecture allows us to easily expand our team of agents to include new members with unique skills needed for financial analysis.

The next big step could be to integrate Large Action Models (LAMs) into the system. LAMs are designed to understand and execute human intentions by combining perception and action \cite{lam_1}. 
The advanced understanding and action capabilities of LAMs have the potential to fully automate the trading process. Based on ElliottAgents analyses, LAMs could automatically perform trades and adapt to rapidly changing market conditions. However, LAMs are currently in the early stages of adoption, making their implementation in such systems challenging.

\subsection{Conclusion}

ElliottAgents represents a significant advancement in the field of financial technology, bridging the gap between traditional technical analysis and AI methodologies. The platform's success in combining the EWP with multi-agent AI systems opens new area for research in algorithmic trading. By demonstrating the effectiveness of this integrated approach, this research contributes to the ongoing evolution of intelligent financial systems.

The proposed system, ElliottAgents, integrates traditional financial analysis methods with AI technologies to enable a deeper and more precise analysis of historical data for accurate future price predictions. This research presents a system design capable of thoroughly analyzing various American stock market companies across different time frames and intervals. While the system primarily focuses on medium to long-term analysis, it can be customized for shorter intervals like 5 minutes. However short team effectiveness is limited due to price swings and the presence of other algorithms for high frequency trading. It's important to note that unforeseen market events, such as rapid crashes or unexpected news, can significantly impact the accuracy of short-term predictions. 

By demonstrating the effectiveness of this integrated approach, this research contributes to the ongoing evolution of intelligent financial systems. While the current system demonstrates the feasibility of our approach, future work will focus on incorporating additional features and refining the system's predictive capabilities. 

\bibliographystyle{apalike}
{\small
\bibliography{main}}

\end{document}